\documentclass[twocolumn,preprintnumbers,amsmath,amssymb]{revtex4}

\usepackage{rotating}
\usepackage{amsmath}
\usepackage{graphicx,epsfig,floatflt,subfigure}
\newcommand\etal{\mbox{\textit{et al.}}}
\begin{document}

\markboth{G. Lamaison, W.J.T. Bos, L. Shao and J.-P. Bertoglio}{Decay of scalar variance in a bounded domain}

\title{\large Decay of scalar variance in isotropic turbulence in a bounded domain}

\author{Guillevic Lamaison\thanks{guillevic.lamaison@ec-lyon.fr},
Wouter Bos\thanks{wouter.bos@ec-lyon.fr},
Liang Shao and Jean-Pierre Bertoglio\\
LMFA, UMR CNRS 5509, Ecole Centrale de Lyon - UCBL - INSA Lyon,\\ 69134 Ecully, France}
\received{2nd August 2006}

\begin{abstract}
{\bf Abstract}. The decay of scalar variance in isotropic turbulence in a bounded domain is investigated. Extending the study of Touil, Bertoglio and Shao [J. Turbul. {\bf 03}:49 2002] to the case of a passive scalar, the effect of the finite size of the domain on the lengthscales of turbulent eddies and scalar structures is studied by truncating the infrared range of the wavenumber spectra. Analytical arguments based on a simple model for the spectral distributions show that the decay exponent for the variance of scalar fluctuations is proportional to the ratio of the Kolmogorov constant to the Corrsin-Obukhov constant. This result is verified by closure calculations in which the Corrsin-Obukhov constant is artificially varied. Large-Eddy Simulations provide support to the results and give an estimation of the value of the decay exponent and of the scalar to velocity time scale ratio.
\bigskip

\end{abstract}

\maketitle

\section{Introduction}

Most turbulent flows on earth are bounded by walls. An important part of theoretical studies of turbulence is however devoted to homogeneous fields, ruling out the existence of boundaries that are present in most situations of practical interest or laboratory experiments. Even in nearly homogeneous flows, such as decaying grid turbulence, the integral lengthscale will grow indefinitely so that, if the turbulence was initially strong enough, this lengthscale will sooner or later become comparable to the size of the domain. Evidently, the understanding of spatially bounded turbulence is of primary importance in a wide range of academic and engineering problems and remains one of the major challenges in turbulence research.

Indeed, the presence of walls adds an enormous complexity to the description of turbulence. Without taking into account the zoology of structures and phenomena associated with the existence of a strong shear near the boundaries present in most wall bounded flows, there is a simple academic case where global predictions can easily be  performed. This is the case of an isotropic turbulence freely evolving in a bounded domain. Following Tennekes and Lumley \cite{Tennekes} it can be  predicted that, after a period of free decay, in which the kinetic energy evolves classically, as soon as the integral lengthscale becomes limited by the size of the domain, the turbulent kinetic energy $k$ decays as $t^{-2}$ (and therefore that the dissipation $\epsilon$ and rms vorticity decay as $t^{-3}$ and  as $t^{-3/2}$ respectively). This prediction was confirmed experimentally by Skrbek and Stalp \cite{Skrbek} in confined superfluid grid turbulence. After the lengthscale saturation occured, they measured a $t^{-3/2}$ decay for the rms vorticity and performed an analysis of the results based on a simple model spectrum with an infrared cut-off. It was already anticipated  by Bertoglio and Jeandel \cite{Berto}, that the scale limitation in spectral studies of turbulence could be roughly taken into account by imposing an infrared cut-off in the energy spectrum, representing the fact that eddies larger than the domain size cannot exist.  Direct numerical simulations (DNS), Large-Eddy Simulations (LES) and two-point closure calculations \cite{Touil2} 
confirmed the result for isotropic turbulence. The effect of a bounded domain on anisotropic turbulence was similarly investigated by Biferale \etal \cite{Biferale} by DNS. The present work continues this line of research by investigating the decay of passive scalar fluctuations in a confined turbulent flow. This effort contributes to a better understanding of turbulent mixing in bounded domains.

As stated above, the decay exponent of the kinetic energy can be estimated by using the relation (Tennekes and Lumley \cite{Tennekes})
\begin{equation}\label{epsu3L}
\epsilon\sim \frac{k^{3/2}}{d}
\end{equation}
with $d$ the size of the domain that determines the integral lengthscale. Assuming  self-similar decay of the turbulent energy spectrum, power law decay can be expected for $k$ and $\epsilon$:
\begin{eqnarray}\label{PLs1}
k\sim(t-t_0)^{-n} \qquad&\qquad \epsilon \sim  n(t-t_0)^{-(n+1)}
\end{eqnarray}
with $t_0$ a virtual origin. Using (\ref{epsu3L}) and the relation
\begin{equation}\label{kt=eps}
k_{,t}=-\epsilon,
\end{equation}
one obtains
\begin{equation}
(t-t_0)^{-(n+1)}\sim d^{-1}(t-t_0)^{-\frac{3}{2}n}
\end{equation}
so that the equality of the exponents leads to the already mentioned result $n=2$. 

The equation for the variance of passive scalar fluctuations $\overline{\theta^2}$ in isotropic turbulence without mean scalar gradients is
\begin{equation}\label{epst}
\frac{1}{2}\overline{\theta^2}_{,t}=-\epsilon_\theta.
\end{equation}
in which $\epsilon_{\theta}$ is the dissipation of scalar fluctuations  $\epsilon_{\theta}$. 

It is tempting to apply the above  reasoning for the kinetic energy to the decay of the passive scalar. The equivalent of (\ref{epsu3L}) for scalar decay is:
\begin{equation}
\epsilon_{\theta}\sim \frac{\overline{\theta^2} k^{1/2}}{d}
\end{equation}
assuming power laws for $k$, $\epsilon_{\theta}$ and $\overline{\theta^2}$ yields:
\begin{equation}
(t-t_0)^{-(n_\theta+1)}\sim d^{-1}(t-t_0)^{-(n_\theta+n/2)}
\end{equation}
which yields by substituting $n=2$ and using the equality of the exponents
\begin{equation}\label{n=n}
-(n_\theta+1)=-(n_\theta+1)
\end{equation}
which is verified for every $n_\theta$. We therefore need to use another method to determine $n_\theta$. This is the purpose of section \ref{secAnalysis}, in which  an analytical study based on a simplified spectral form of the energy and scalar spectra is performed that allows to derive analytical expressions for the scalar decay. These expressions are then validated by two-point closure computations of the eddy-damped quasi-normal Markovian (EDQNM) type and confirmed by Large-Eddy Simulations in section \ref{secEDQNM}. All through the paper the Schmidt number will be assumed unity.

\section{Analysis and scalings \label{secAnalysis}}


\begin{figure} 
\setlength{\unitlength}{1.\textwidth}
\includegraphics[width=0.45\unitlength]{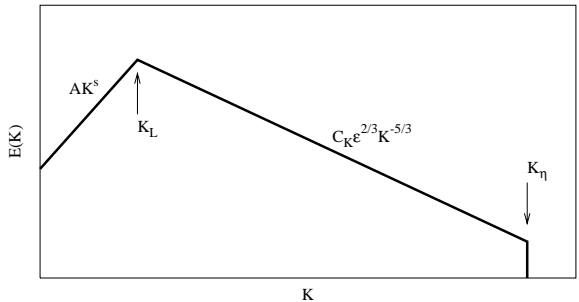}
\caption{\textit{Model spectrum used to predict selfsimilar decay of turbulence. \label{FigTP}}}
\end{figure}

For the purpose of clarity some results for the velocity field are recalled, before starting the discussion of the passive scalar. The decay of kinetic energy in an unbounded isotropic turbulent field can be roughly predicted by considering the simple model spectrum (c.f. Comte-Bellot and Corrsin \cite{CBC}),
\begin{equation} 
E(K) = 
\begin{cases}\label{modspec}
AK^s & \textrm{for}~  K<K_{L}\\
C_K\epsilon^{2/3}K^{-5/3} & \textrm{for}~  K_L\le K \le K_\eta \\
0 & \textrm{for}~  K > K_\eta
\end{cases}
\end{equation}
where $C_K$ is the Kolmogorov constant. This spectrum corresponds to the sketch in figure \ref{FigTP}. It will be assumed in the following that $K_\eta\gg K_L$. $E(K)$ is related to the kinetic energy $k$ by
\begin{eqnarray}\label{quants}
k=\int_0^{\infty}E(K)dK
\end{eqnarray}
Assuming power laws for $k$ and $\epsilon$ (\ref{PLs1}), and that $A$ is  constant during the time evolution, the following decay exponent is obtained 
\begin{eqnarray}\label{ns}
n=\frac{2(s+1)}{s+3}.
\end{eqnarray}
Note that the constancy of $A$ is a reasonable assumption for $s<4$ (see Lesieur and Schertzer \cite{Lesieur78}). A study of the time evolution of $A$ for $s=4$ can be found for example in Chasnov \cite{Chasnov93}, but is outside the scope of the present paper. The present work focuses on the behavior of turbulence once the growth of the integral scales is limited by the domain size. As shown in Skrbek and Stalp \cite{Skrbek} or Touil \etal \cite{Touil2}, as soon as the integral scale becomes comparable to the domain size, the decay exponent tends to $2$, a value larger than before saturation ($n=10/7$ for $s=4$ for example). This exponent $2$ can be deduced assuming an energy spectrum with a sharp infrared cut-off, of the form
\begin{equation} 
E(K) = 
\begin{cases}\label{modspec0}
0 & \textrm{for}~  K<K_{inf}\\
C_K\epsilon^{2/3}K^{-5/3} & \textrm{for}~  K_{inf}\le K \le K_\eta \\
0 & \textrm{for}~  K > K_\eta
\end{cases}
\end{equation}
i.e. an inertial range extending from the wavenumber $K_{inf}=2\pi/d$, with $d$ the domain size, to $K_\eta$. 

Integrating (\ref{modspec0}) gives in the limit of an infinite Reynolds number:
\begin{equation}\label{ggg}
k=\frac{3}{2}C_K\epsilon^{2/3}K_{inf}^{-2/3}.
\end{equation}
Assuming power law decay for $\epsilon$ (\ref{PLs1}) and using (\ref{kt=eps}) one can express $\epsilon$ as:
\begin{eqnarray}\label{expeps}
\epsilon=\frac{C_K^3}{K_{inf}^2}\frac{(n+1)^3}{(t-t_0)^3}\\
	\sim(t-t_0)^{-(n+1)}\nonumber
\end{eqnarray}
so that the equality between time exponents in the balance implies that $n=2$.

Let us now come back to the central issue of this work, the decay of the variance of passive scalar fluctuations. By analogy with (\ref{modspec}) the scalar spectrum can be assumed to have the form:
\begin{equation} 
E_{\theta}(K) = 
\begin{cases}\label{modspecT}
A_{\theta}K^{s'} & \textrm{for}~  K<K_{\theta}\\
C_{CO}\epsilon^{-1/3}\epsilon_{\theta}K^{-5/3} & \textrm{for}~  K_{\theta}\le K \le K_\eta \\
0 & \textrm{for}~  K > K_\eta
\end{cases}
\end{equation}
with $C_{CO}$ the Corrsin-Obukhov constant. The scalar variance is related to the spectrum by
\begin{eqnarray}\label{quants}
\frac{1}{2}\overline{\theta^2}=\int_0^{\infty}E_{\theta}(K)dK.
\end{eqnarray}
Let us assume that $\overline{\theta^2}$ decays as a power law:
\begin{eqnarray}\label{PLs2}
\overline{\theta^2}\sim(t-t_0)^{-n_\theta} 
\end{eqnarray}
then $\epsilon_{\theta}$ decays as
\begin{eqnarray}\label{PLs3}
\epsilon_\theta \sim  n_\theta(t-t_0)^{-(n_\theta+1)}.
\end{eqnarray}
This is not a trivial assumption and it will be checked a posteriori by closure computations. The decay of $\overline{\theta^2}$ depends not only on its spectral distribution, but also on the injection scale of the scalar fluctuations \cite{CorrsinJAS,Warhaft78,Ristorcelli}. 
In the present work we  consider the case in which the initial scales of the scalar field and the velocity field are comparable, $K_L\approx K_\theta$. Within this framework, power law decay can be possible \cite{Lesieur}, and an expression similar to (\ref{ns}) can be derived for the  scalar variance. The resulting expression (Herring \etal \cite{Herring}) is: 
\begin{eqnarray}\label{nsT}
n_\theta=\frac{2(s'+1)}{s+3}.
\end{eqnarray}
An important quantity directly related to the decay of scalar variance and appearing in most engineering models, is the velocity to scalar time scale ratio, defined by:
\begin{equation}\label{rrr}
r=\frac{2k\epsilon_{\theta}}{\overline{\theta^2}\epsilon}.
\end{equation}
The assumption of power law decay of the integral quantities  (\ref{PLs1}) and (\ref{PLs2}) yields immediately the result $r=n_\theta/n$, which gives for the present case
\begin{equation}\label{rrr2}
r=\frac{s'+1}{s+1}.
\end{equation}
A more elaborate expression, depending on the Schmidt number, was recently proposed by Ristorcelli \cite{Ristorcelli}. 

To investigate the effect of the limitation of the turbulent and scalar lengthscales when the integral lengthscales become comparable to the size of the domain, we still assume that the energy spectrum is given by (\ref{modspec0}) and we furthermore postulate that the scalar spectrum has the form
\begin{equation} 
E_{\theta}(K) = 
\begin{cases}\label{modspecT0}
0 & \textrm{for}~  K<K_{inf}\\
C_{CO}\epsilon^{-1/3}\epsilon_{\theta}K^{-5/3} & \textrm{for}~  K_{inf}\le K \le K_\eta \\
0 & \textrm{for}~  K > K_\eta.
\end{cases}
\end{equation}
Using expression (\ref{expeps}) for the dissipation in (\ref{modspecT0}) and integrating to obtain $\overline{\theta^2}$:
\begin{equation}\label{ggg}
\frac{1}{2}\overline{\theta^2}=\frac{3}{2}C_{CO}\epsilon_{\theta}\epsilon^{-1/3}K_{inf}^{-2/3},
\end{equation}
and using (\ref{PLs2}) together with (\ref{epst}) gives 
\begin{equation}\label{prefs}
-n_\theta\frac{C_{CO}}{2C_K}(t-t_0)^{-(n_\theta+1)}=-(t-t_0)^{-(n_\theta+1)}
\end{equation}
so that equality of the exponents, leads again to expression  (\ref{n=n}) which is satisfied whatever the value of $n_\theta$. To deduce an estimation of $n_\theta$ we must therefore use the equality between the prefactors in (\ref{prefs}). It is found that 
\begin{equation}\label{eqn}
n_\theta=2\frac{C_k}{C_{CO}}.
\end{equation}
The time scale ratio directly follows
\begin{equation}\label{eqr}
r={n_\theta}/2
\end{equation}
It must be emphasized that the situation is different from what is found for the kinetic energy spectrum, or for the scalar freely decaying in an unbounded domain, where the values of the decay exponents are found by simply using the relation between the exponents. The fact that here use is made of a relation between the prefactors, leads to the rather untypical situation where the Corrsin-Obukhov and Kolmogorov constants expicitly appear in the expression of the decay exponents. This is for example in contrast with what is found for the velocity field where $n$ is not a function of the Kolmogorov constant, but is entirely determined by the power law exponent $s$ of the spectrum in the low wavenumber region. 

The relations (\ref{eqn}) and (\ref{eqr}) stress the importance of an precise knowledge of $C_K$ and $C_{CO}$. A review aimed at determining the precise value of $C_{CO}$ by a compilation of numerous experimental results  was performed by Sreenivasan \cite{Sreenivasan}. In this work the value of $C_{CO}'$, the constant intervening in the one-dimensional scalar spectrum was found to be approximately $0.3<C_{CO}'<0.6$. $C_{CO}'$ is related to $C_{CO}$ by the relation $C_{CO}=(5/3)C_{CO}'$ which gives $0.5<C_{CO}<1.0$ according to the experiments.

Using EDQNM and LES it will be investigated in the following section if the predictions of $r$ and $n_\theta$ are right.

\section{Closure and Large-Eddy Simulation \label{secEDQNM}}

\begin{figure}[!ht]
\begin{center}
\includegraphics[scale=0.6,angle=0]{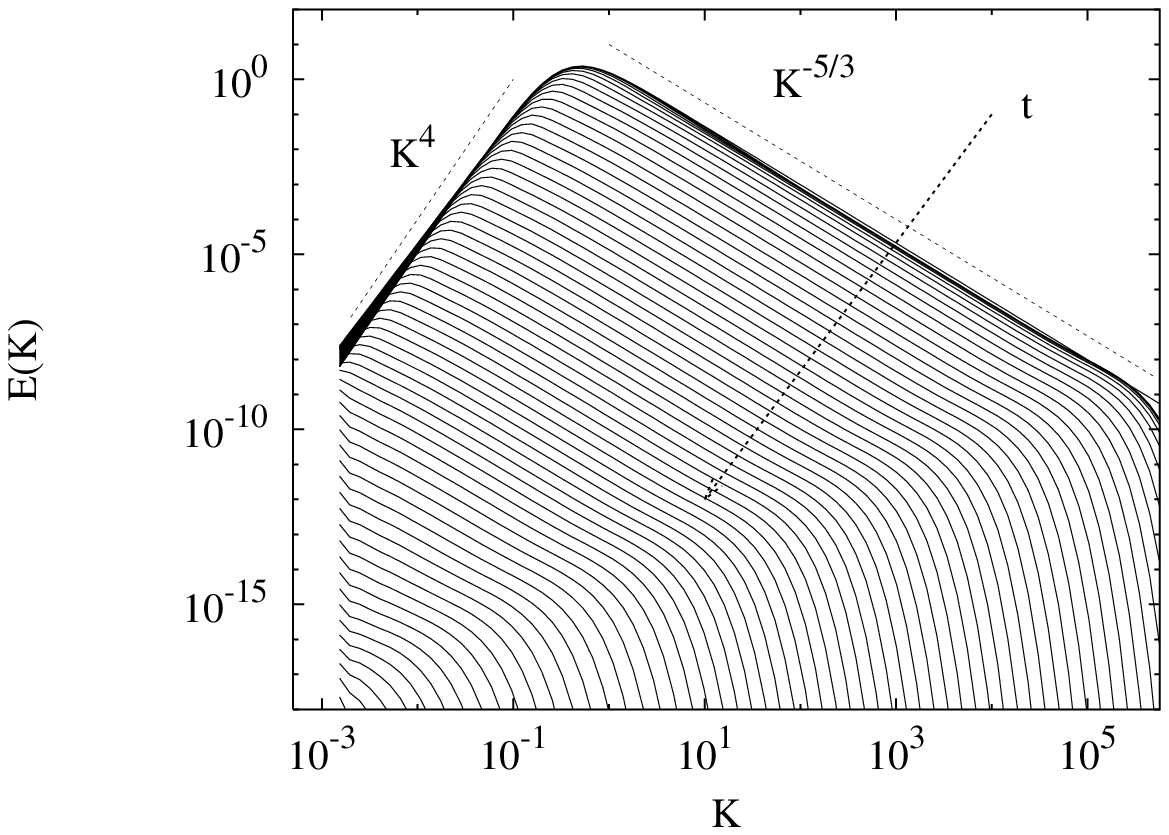}
\includegraphics[scale=0.6,angle=0]{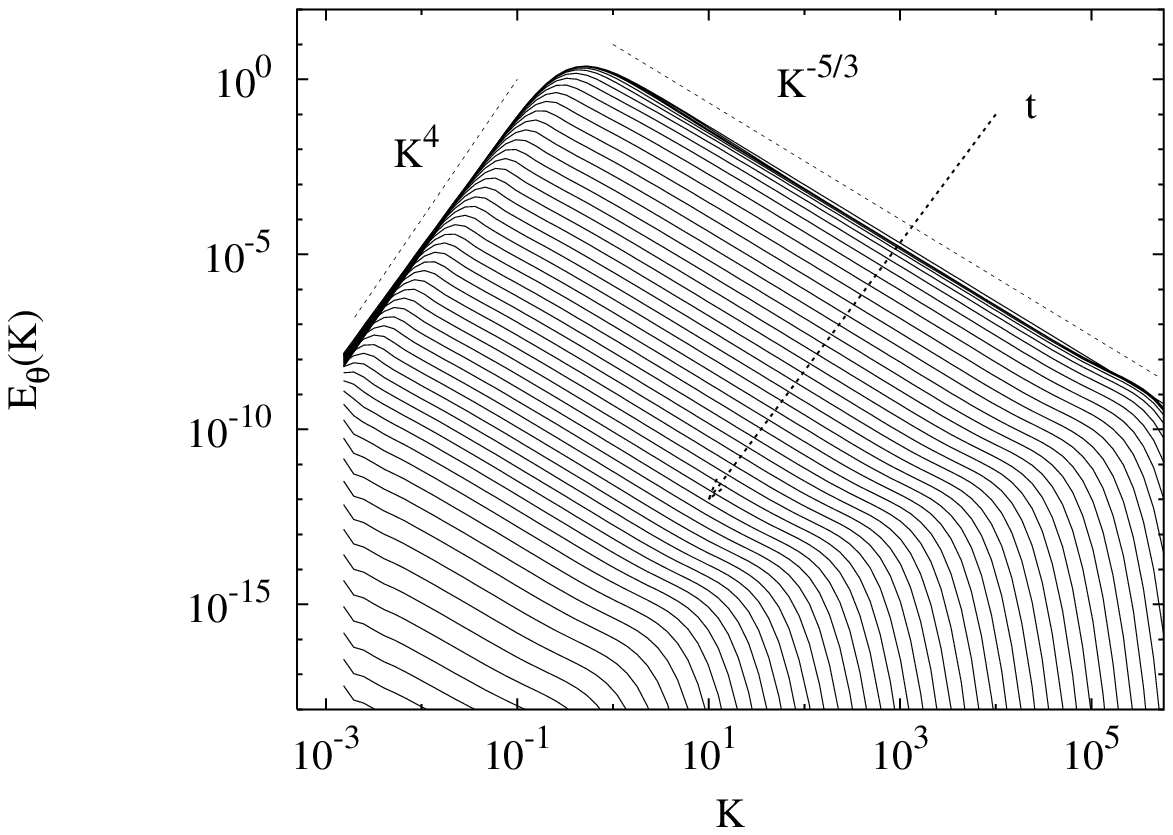}
\caption{\textit{Time evolution of the kinetic energy spectrum (top) and scalar variance spectrum (bottom)}}
\label{fig:1a}
\end{center}
\end{figure}

\subsection{EDQNM}
In this paper computations are performed within the framework of the eddy-damped quasi-normal Markovian (EDQNM) closure, as proposed for isotropic turbulence by Orszag \cite{Orszag} and Leith \cite{Leith2}. The isotropic scalar formulation used here was first proposed by Vignon \etal \cite{VignonCras,VignonPOF} and Herring \etal \cite{Herring}. The evolution equations for the kinetic energy spectrum of an isotropic turbulence and for the passive scalar spectrum are respectively:
\begin{equation}
\frac{\partial}{\partial t} E(K) = -2 \nu K^2  E(K) +  T(K)
\end{equation}
\begin{equation}
\frac{\partial}{\partial t} E_{\theta}(K) = -2 \kappa K^2  E_{\theta}(K) +  T_{\theta}(K)
\end{equation}
Time dependence is omitted for notational simplicity. The viscosity $\nu$ is equal to the diffusivity $\kappa$ for the case of unity Schmidt number, which we consider here. The non-linear transfer terms $ T(K)$ and $T_{\theta}(K)$ are expressed using the classical
EDQNM formulation for isotropic turbulence and scalar field:
\begin{eqnarray}
T(K)=\int_{\Delta(K)} \theta_{KPQ}(t)\frac{xy+z^{3}}{Q}
E(Q) \times \nonumber\\
\bigg\{ K^{2}E(P)- P^{2}E(K) \bigg\} dPdQ
\end{eqnarray}
\begin{eqnarray}
T_{\theta}(K)=\int_{\Delta(K)} \theta_{KPQ}^{\theta}(t)\frac{1-z^{2}}{Q^{3}}
E(Q)
\times \nonumber\\
\bigg\{ K^{3}E_{\theta}(P)- KP^{2}E_{\theta}(K) \bigg\} dPdQ
\end{eqnarray}
in which $\Delta$ denotes the domain such that the three wave vectors ${\bm{K}, \bm{P}, \bm{Q}}$ form a triangle and $x$, $y$ and $z$ are the cosines of
the angles respectively opposite to ${\bm{K}, \bm{P}, \bm{Q}}$ in this triangle. The EDQNM characteristic times are
given by 
\begin{equation}
\theta_{KPQ}(t)=\frac{1-e^{-(\nu (K^{2}+ P^{2}+ Q^{2})+\eta(K)+\eta(P)+\eta(Q))t}}{\nu (K^{2}+ P^{2}+ Q^{2})+\eta(K)+\eta(P)+\eta(Q)}
\end{equation}
\begin{equation}
\theta_{KPQ}^{\theta}(t)=\frac{1-e^{-(\kappa (K^{2}+P^2)+\nu Q^{2}+ \eta'(K)+\eta'(P)+\eta''(Q))t}}{\kappa (K^{2}+P^2)+\nu Q^{2}+\eta'(K)+\eta'(P)+\eta''(Q)}
\end{equation}
in which the damping coeficients are expressed using the classical forms:
\begin{eqnarray}
\eta(K)= \lambda \sqrt{\int_{0}^{K}R^{2}E(R)dR}\nonumber\\
\eta'(K)= \lambda_{1} \sqrt{\int_{0}^{K}R^{2}E(R)dR} \nonumber\\
\eta''(K)= \lambda_{2} \sqrt{\int_{0}^{K}R^{2}E(R)dR}.
\end{eqnarray}
Three constants have to be fixed. For $\lambda$ we use the classical value
$\lambda=0.355$. The values for $\lambda_{1}$ and $\lambda_{2}$ require more attention. We follow Herring \etal \cite{Herring} and choose a zero value for $\lambda_{1}$ for compatability with the Lagrangian History Direct Interaction Approximation of Kraichnan \cite{Kraichnan65}. The value of $\lambda_{2}$ is not fixed and is related to the Corrsin-Obukhov constant. We will first use the classical value (see Lesieur \cite{Lesieur} and Herring \etal \cite{Herring}) $\lambda_{2}=1.3$. The results will be analyzed in the following section. Then, we will take advantage of the fact that varying $\lambda_{2}$ provides a simple and efficient way to vary the Corrsin-Obukhov constant in the model and we will use this degree of freedom to check relations (\ref{eqn}) and (\ref{eqr}). 

The computational domain ranges from $K_{inf}$ (low-wavenumber or infrared cut-off, related to the size $d$ of the bounded domain by $K_{inf}=2 \pi /d$) to $4K_{\eta}$, $K_{\eta}$ being the Kolmogorov wavenumber). The resolution is approximately 14 wavenumbers per decade. The initial spectrum is a \emph{von Karman} spectrum \cite{Hinze}. The energy-containing range is characterized by wavenumber $K_L$, the wavenumber at which $E(K)$ has its maximum. At time $t=0$, initial conditions are  such that $K_L$ is greater than $K_{inf}$. The Taylor-scale Reynolds number evaluated at the time when $K_L\approx K_{inf}$ is approximately $10^5$. This high value for the Reynolds number is chosen to allow a precise determination of the Kolmogorov and Corrsin-Obukhov constants. 

\subsection{Large-Eddy Simulation \label{secLES}}

The code used for the LES computations is a classical pseudo-spectral code with fourth-order Runge-Kutta time integration scheme. The resolution is $128^3$ grid points. As initial spectrum, a \emph{von Karman} spectrum is used which behaves as $K^4$ at small $K$ and as $K^{-5/3}$ for large $K$. The CZZS dynamic model \cite{Cui,Shao04} is used to model the subgrid stress and scalar flux. 

\section{Results}

In Figure \ref{fig:1a} EDQNM results are shown for the time evolution of the kinetic energy spectrum and the scalar variance spectrum. It can be observed that both spectra display a clear $K^{-5/3}$ inertial range. The inertial ranges are observed before saturation and are still present after.

\begin{figure}[!ht]
\begin{center}
\includegraphics[scale=0.6,angle=0]{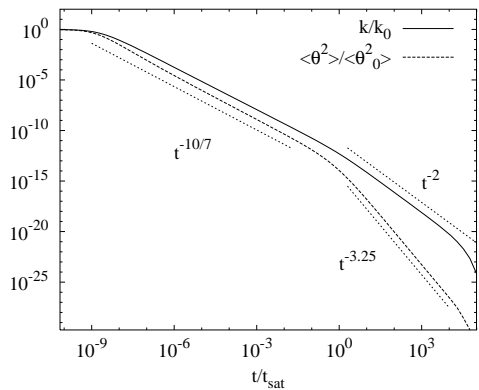}
\caption{\textit{Decay of kinetic energy and scalar fluctuations. EDQNM computations results}}
\label{fig:0a}
\end{center}
\end{figure}

\begin{figure}[!ht]
\begin{center}
\includegraphics[scale=0.6,angle=0]{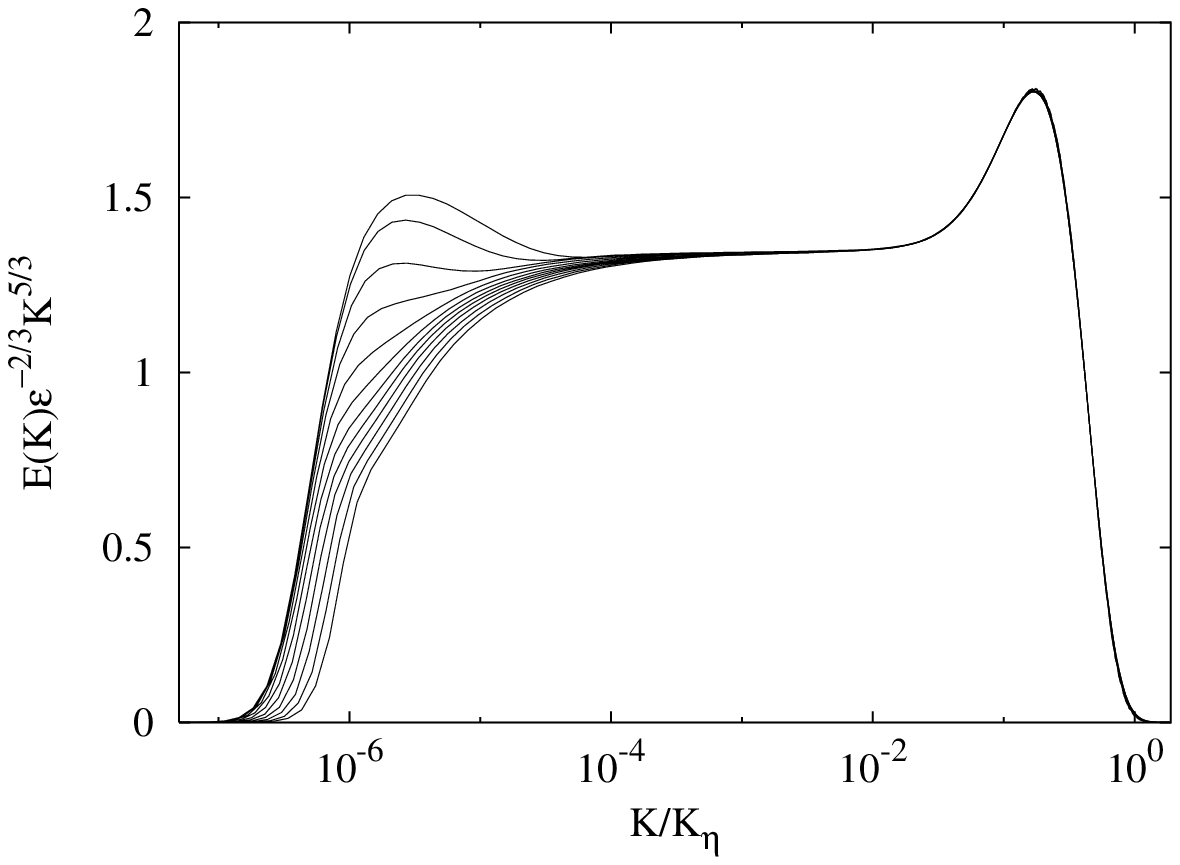}
\includegraphics[scale=0.6,angle=0]{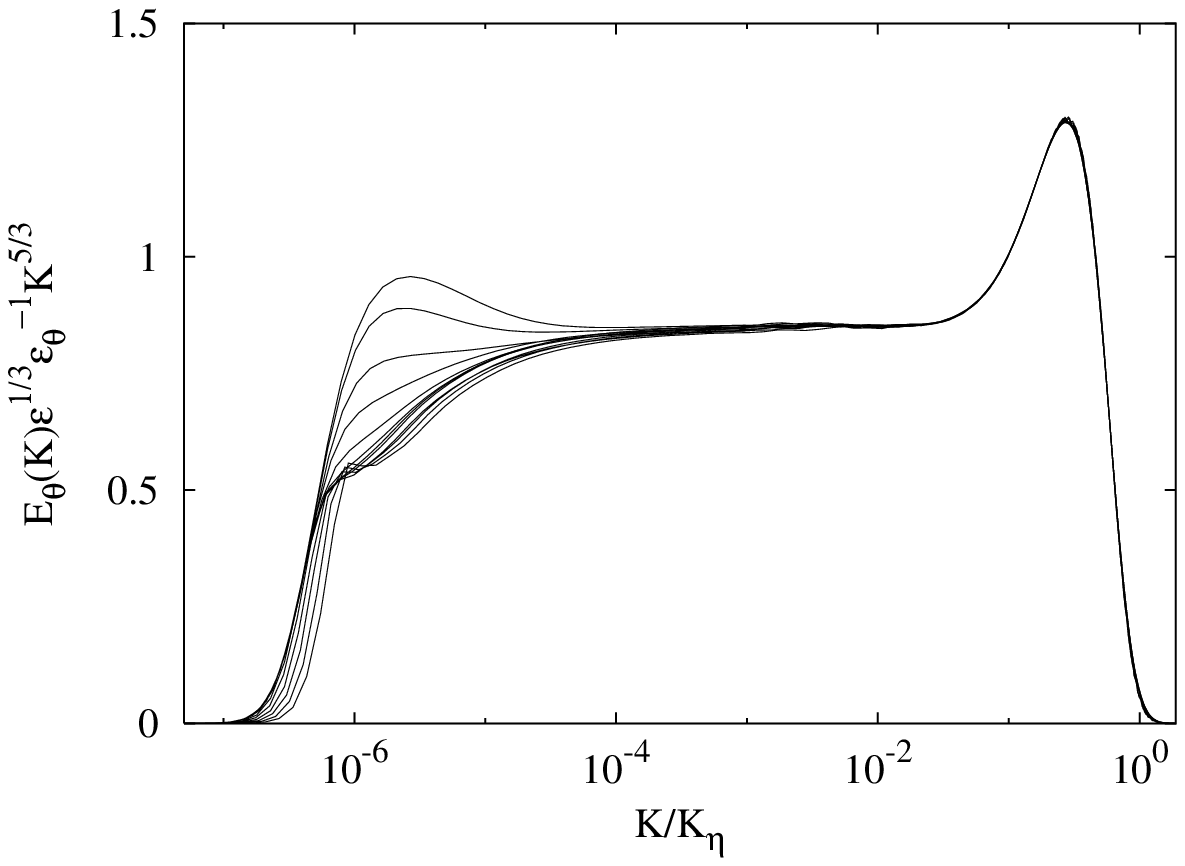}
\caption{\textit{Compensated kinetic energy spectrum (left) and scalar variance spectrum (right)}}
\label{fig:3a}
\end{center}
\end{figure}


These results were obtained with the classical values of the constants ($\lambda=0.355$, $\lambda_{1}=0$ and $\lambda_{2}=1.3$). The kinetic energy and scalar variance are shown in figure \ref{fig:0a}. It is observed that after a short time the kinetic energy decays following a power law with an exponent close to the classical value $10/7$. As soon as the large scales become saturated by the presence of the infrared cut-off, at $t_{sat}$, the power law exponent changes and takes a value of $2$ as predicted in the preceding sections. At very long times, the behavior changes, corresponding to a final period of viscous decay. It is observed in figure \ref{fig:0a}, that the scalar variance also decays following a power law with an exponent close to $10/7$. At longer times, after saturation has occurred, the decay is found to follow a $t^{-3.25}$ power law. The Kolmogorov and Corrsin-Obukhov constants are determined by carefully evaluating the compensated spectra. Figure \ref{fig:3a} shows the compensated spectra $E(K)$ and $E_\theta(K)$. The Kolmogorov constant is $C_K\approx 1.35$, a slightly low value. The Corrsin-Obukhov constant is then approximately $0.85$. The ratio $2C_k/C_{CO}$ yields $3.2$ which is close to the value of $n_\theta$ observed in figure \ref{fig:0a}.

\begin{figure}[!ht]
\begin{center}
\includegraphics[scale=0.6,angle=0]{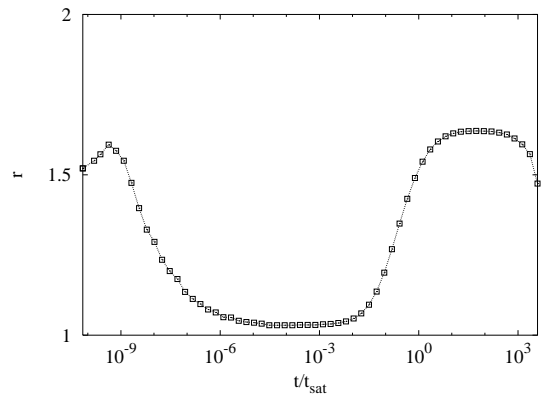} 
\caption{\textit{Time evolution of the velocity to scalar time scale ratio}}
\label{fig:5}
\end{center}
\end{figure}

Figure \ref{fig:5} shows the time evolution of the velocity to scalar time scale ratio. It is observed that after a transient, $r$ goes to a constant value close to $1$ corresponding to the freely decaying period where both $k$ and $\overline{\theta^2}$ evolve as $t^{-10/7}$. At longer times, after saturation has occurred, a second plateau is observed with a value close to $1.6$. This value is in agreement with the analysis in section  \ref{secAnalysis} that predicts that $r=C_k/C_{CO}$.

\begin{figure}[!ht]
\begin{center}
\includegraphics[scale=0.6,angle=0]{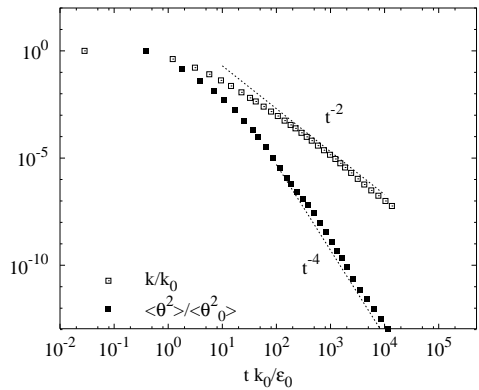}
\caption{\textit{Decay of kinetic energy and scalar fluctuations. LES results}}
\label{fig:0b}
\end{center}
\end{figure}

\begin{figure}[!ht]
\begin{center} 
\includegraphics[scale=0.6,angle=0]{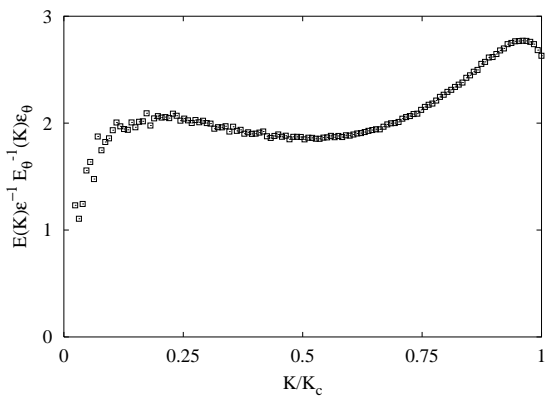}
\caption{\textit{Ratio $C_K/C_{CO}$ from the Large-Eddy Simulations}}
\label{FigLESC}
\end{center}
\end{figure}

In figure \ref{fig:0b}  the kinetic energy and scalar variance from the Large-Eddy Simulation are shown. For the saturated decay a power law exponent for the energy close to $2$ is observed in agreement with the results of Touil \etal \cite{Touil2}. The scalar variance decays with an exponent close to $4$. This corresponds according to (\ref{eqn}) to a ratio $C_K/C_{CO}$ close to $2$. This is well verified as can be oberved in figure \ref{FigLESC} where the ratio $C_K/C_{CO}$ is plotted as a function of $K/K_c$ with $K_c$ the cut-off wavenumber. 

\begin{figure}[!ht]
\begin{center}
\includegraphics[scale=0.6,angle=0]{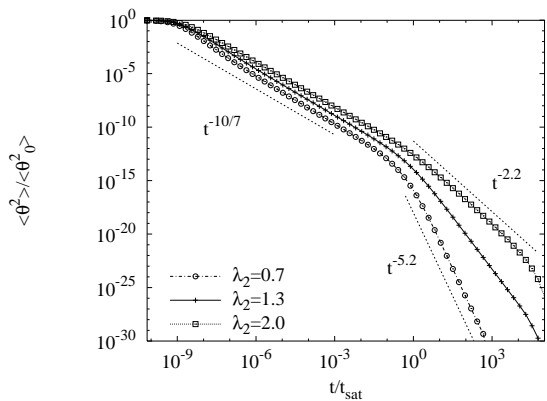}
\caption{\textit{Decay of passive scalar variance for varying $\lambda_2$}}
\label{fig:2a}
\end{center}
\end{figure}

To check the relations (\ref{eqn}) and (\ref{eqr}) in more detail $C_{CO}$ is now artificially varied by changing the value of $\lambda_{2}$ in the range [0.7-2.0]. By examining the compensated spectra it was observed that this corresponded to a Corrsin-Obukhov constant that takes values in the range [0.46-1.28]. 
The time evolutions of $\overline{\theta^2}$ are shown in figure \ref{fig:2a} for three different values of $\lambda_2$. Clear power laws are observed. The power law exponent is affected by the variation of $\lambda_2$. 

\begin{figure}[!ht]
\begin{center}
\includegraphics[scale=0.6,angle=0]{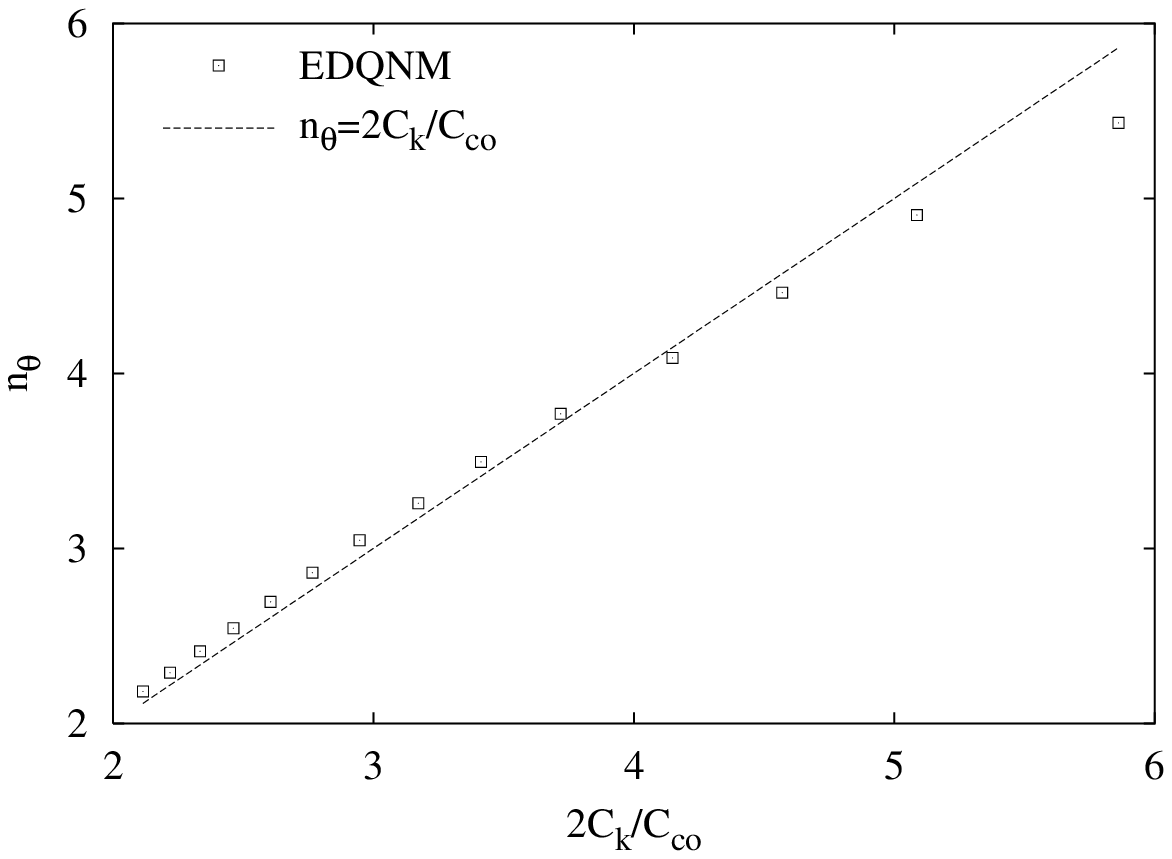}
\includegraphics[scale=0.6,angle=0]{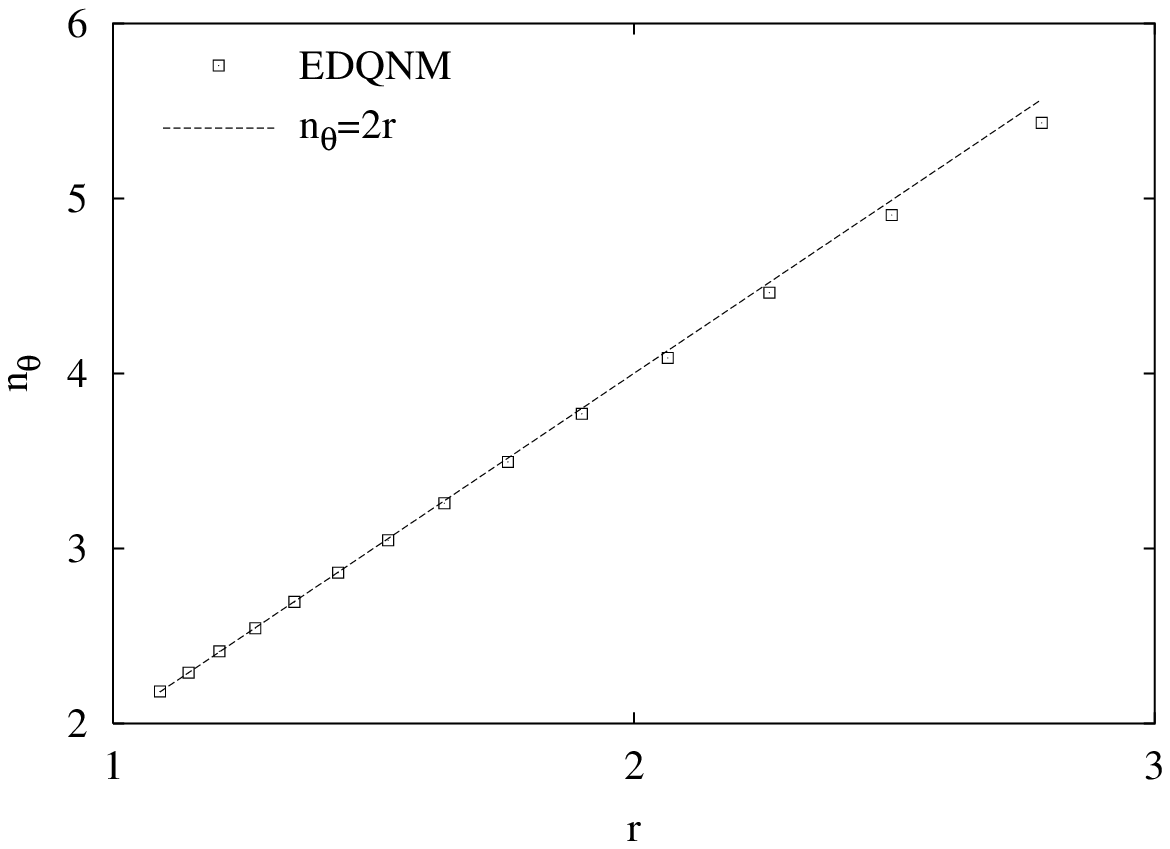}
\caption{\textit{Scalar decay exponent $n_\theta$ as a function of the ratio of the Kolmogorov $C_k$ and Corrsin-Obukhov constant $C_{CO}$ (top). Scalar decay exponent as a function of the time scale ratio $r$ (bottom)}}
\label{fig:4a}
\end{center}
\end{figure}

In figure \ref{fig:4a} we test the results (\ref{eqn}) and (\ref{eqr}) against our computations. Good agreement is found between the results of the closure and the analytical results (\ref{eqn}) and (\ref{eqr}).

\section{Conclusion}

Analytical expressions were derived for the decay exponent of the scalar variance and the velocity to scalar time scale ratio for a passive scalar decaying in bounded isotropic turbulence. These expressions depend on the Kolmogorov and Corrsin-Obukhov constants. It is rather uncommon in turbulence theory that a decay exponent depends on the values of these inertial range constants. The proposed relations were tested against EDQNM computations in which the Corrsin-Obukhov constant was artificially varied. Good agreement was observed and clear power law decay was observed for time evolution of the scalar variance. The values for the decay exponent and Kolmogorov and Corrsin-Obukhov constants obtained by Large-Eddy Simulation are consistent with these results. The decay exponent is shown to be approximately $4$ in the LES, a significantly larger value than the exponent for the turbulent kinetic energy which in this case is close to $2$. The present study shows the importance of a precise knowledge of spectral constants as their values can directly determine the temporal behavior of integral quantities. 

An experimental verification of the results presented in this paper would probably be difficult to achieve, as in usual laboratory experiments, the Reynolds number is too low for the wall bounded regime to be observed before the final viscous decay occurs. To overcome this difficulty, Skrbek and Stalp \cite{Skrbek} performed measurements in Helium superfluid. They were able to measure the $-3/2$ decay exponent for the rms vorticity that corresponds to the bounded domain regime, but no data on scalar fluctuations were reported in their experimental study.

\section*{Acknowledgments}
The authors would like to thank H. Touil for helpful discussions and M. Lesieur for an essential routine used in the EDQNM computations. 

\label{lastpage}


\end{document}